# Promoting Efficient Infrastructure Competition through Vectoring and Enlarged DSL Bandwidth

Francesco Vatalaro, Franco Mazzenga

University of Roma "Tor Vergata", Rome, Italy

Romeo Giuliano (*)

Guglielmo Marconi University, Rome, Italy

(*) Contact: r.giuliano@unimarconi.it

**Abstract**

The most effective way to promote Next Generation Access coverage and ultra-broadband penetration is through infrastructure competition. In Countries having well-developed cable networks this market condition preexists. Where CaTV is limited or null, one way to surrogate its powerful procompetitive effect is through the VDSL2/FttC architecture stimulating competition among operators. This has been the approach finally selected in Italy to promote the 30 Mbit/s Digital Agenda for Europe objective for year 2020 (DAE-2020). However, some technical intricacies make hard using "vectoring" while the NRA imposes a regulatory remedy known as SLU (Sub Loop Unbundling). In particular, as we verified with extensive computer calculations on the Italian copper network, in a Multi-cabinet architecture the current VDSL2 17a standard profile having 17.6 MHz bandwidth is unable to provide the 100 Mbit/s DAE-2020 policy target. This paper introduces one possible solution to the problem of providing more than 100 Mbit/s coverage to a large population extent (e.g., 70%), useful for a Country, such as Italy, characterized by a set of severe constraints – i.e., geographic, demographic and urbanistic constraints. In the paper, we describe SBV (Sub-band Vectoring) solutions to boost the data-rate under multi-operator conditions, while the NRA encourages infrastructure competition. The SBV solution allow sharing bandwidth, typically between two or more operators independently implementing vectoring. SBV allows the simultaneous use of UBB data transmission of co-sited operators not coordinated nor synchronized with each other. The paper shows results of coverage with SBV in the presence of two and three operators at the same street cabinet using Italy as a case study.

---

# 1. Introduction

When in 2010 ITU-T published the G.993.5 Recommendation [1], also known as "vectoring", the Next Generation Access (NGA) rollout landscape started changing in several Countries. Incumbent operators (IOs) can now deploy vectoring in the field as an enhancement to VDSL2 systems, which allows meeting the UBB (ultra-broadband) 100 Mbit/s objective in existing network scenarios at very limited capital investment cost through FttC (Fiber-to-the-Cabinet) architectures. This deployment also simplifies network operations with respect to the need to run in parallel for several years the legacy copper network and a new FttDp/B/H (Fiber-to-the-Distribution point/Building/Home) network.

Meanwhile, the FttDp/B/H rollout experience was convincing operators that fiber demand and costs were worse than generally expected even in some urban scenarios, while copper performance was exhibiting capabilities to continue improving. Therefore, there was no real urgency switching off copper networks in a short time. In most of those Countries still in mid-stream in the development of optical network access, all this produced a change of viewpoint from a revolutionary copper-to-fiber strategy to a safer and cheaper evolutionary approach to FttH going, where possible, through a FttC intermediate stage.

Even when deploying the fiber access from scratch appears reasonable, a feeling of uncertainty may loom. In the U.S., the Google Fiber project is an interesting and instructive example [2]. The lesson learnt is that universal fiber coverage can still be too premature as a short-term target. This is so, even if sector regulation is light or null (*forbearance*), aerial fiber can be mostly deployed, fees for rights-of-way access are eliminated, and a not negligible fraction of users (5% to 25%) manifest their firm interest in the service (e.g., by providing a down-payment in advance).

In Europe, the situation is quite different as access regulation is generally stronger and focused on differently balancing short-term customer welfare with infrastructure investment objectives. Through regulation at national level and, in some specific cases, even introducing geographic markets and differentiated remedies, regulatory measures can be fine-tuned in Europe. Regulation is generally less demanding in Countries where IOs face CableCo's through vibrant infrastructure competition. In a sense, in such Countries the push on investments doubles. On the one side, it is vital for IOs to follow with investments on their networks the performance improvements of the EuroDOCSIS CaTV standards. On the other side, National Regulatory Authorities (NRAs) are prone to leave market actors more free to compete by weakening some *ex ante* obligations.

Despite this territorially flexible approach, ultra-broadband coverage is still very uneven in Europe.[1] In fact, in those European Countries where competition with CaTV is limited or null, generally NGA advances slowly. This is so, because infrastructure competition is the most virtuous approach to get fast network deployment and service adoption, as academic econometric studies carried out in Europe and elsewhere highlight [3-5], and reports from consulting firms confirm [6].

In Italy, where CaTV is absent, NGA is lagging behind, although recent signs are positive and show a trend to partly recovering the delay. Internationally, Italy is an especially tough case study, due to several concurrent severe constraints. The Country's geography is challenging, due to ample mountainous areas, proportionally not so much present elsewhere (e.g., Germany). Population distribution is fairly even, contrary to some other large European Countries (e.g., France, Spain). Structure of historical centers – not only large cities but also a plethora of medium and small towns – makes fiber installation costs generally high, due to difficult interventions on street pavements, as well as indoors cabling. Italy only has a distinct, but important, advantage: copper network is generally good, since it was largely restructured around 1990. It is also short, especially in the so-called secondary network (Cabinet-to-Buildings). Secondary network paths median length is around 200-250 $m$ nationwide (Figure 1). In this paper, we often refer to Italy because of its challenging conditions in fiber deployment.

---

[1] The difference in NGA coverage between top-ranked Countries and those lagging behind is presently more than 60%.

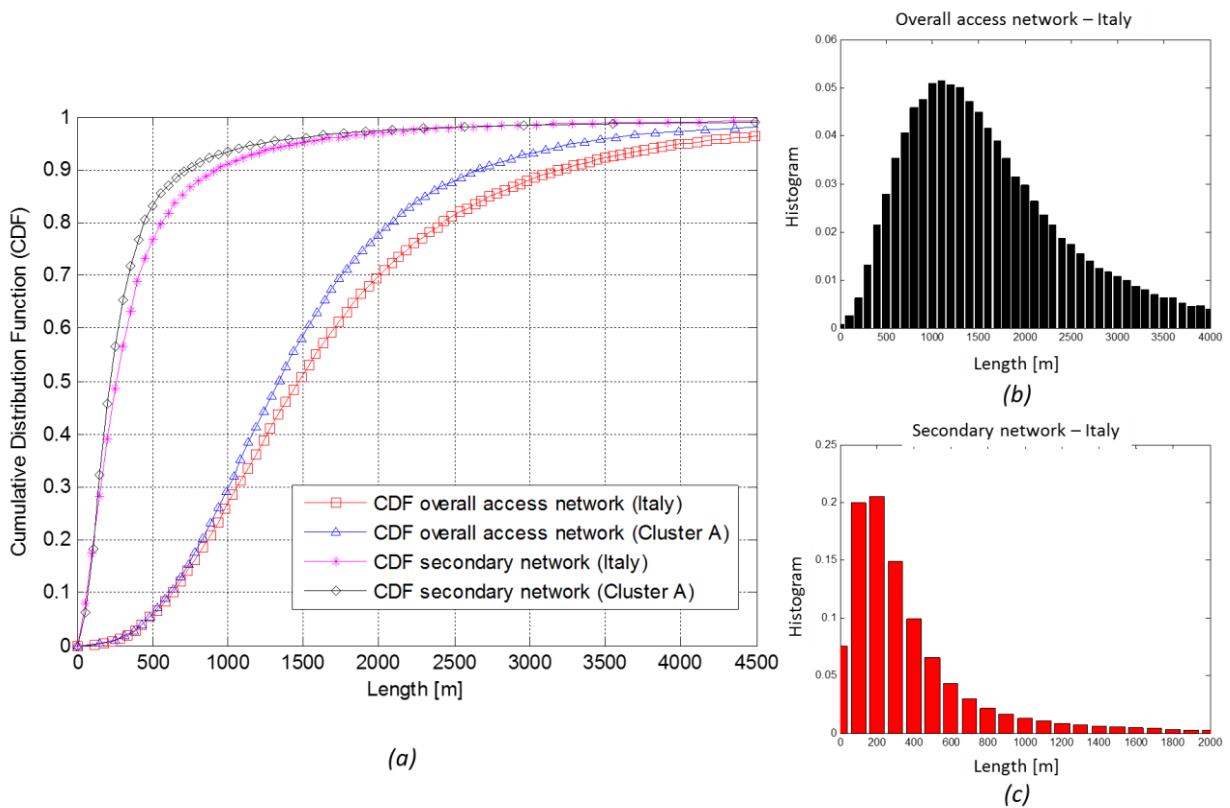

*Figure 1: Statistics of the Italian copper network: (a) cumulative distribution functions for the secondary network and for the overall access network; (b) histogram of the primary network lengths; (c) histogram of the secondary network lengths. (Source: Our elaborations of Italian network data)*

AGCOM, the Italian NRA, has chosen to promote infrastructure competition imposing a set of resolutions [7] hinged in Sub Loop Unbundling (SLU) at the street cabinet (CAB) as a remedy able to promote NGA. This stimulated operators to develop a so-called Multi-CAB architecture (Figure 2), which rapidly found application in several large and medium size cities, with some 50,000 IO's cabinets already equipped all over Italy (about one third).[2]

---

[2] A regulated price mechanism had a powerful effect as incentive for one large infrastructured AO. While AGCOM fixed the ULL price around € 8, the SLU price was about € 3. This price difference was a main driver for said AO to carry all ULL clients to a CAB, irrespective of the ADSL2+ or VDSL2 service sold. Consequently, that AO also adopted commercial policies in order to push clients to upgrade to VDSL2 30 Mbit/s service adoption.

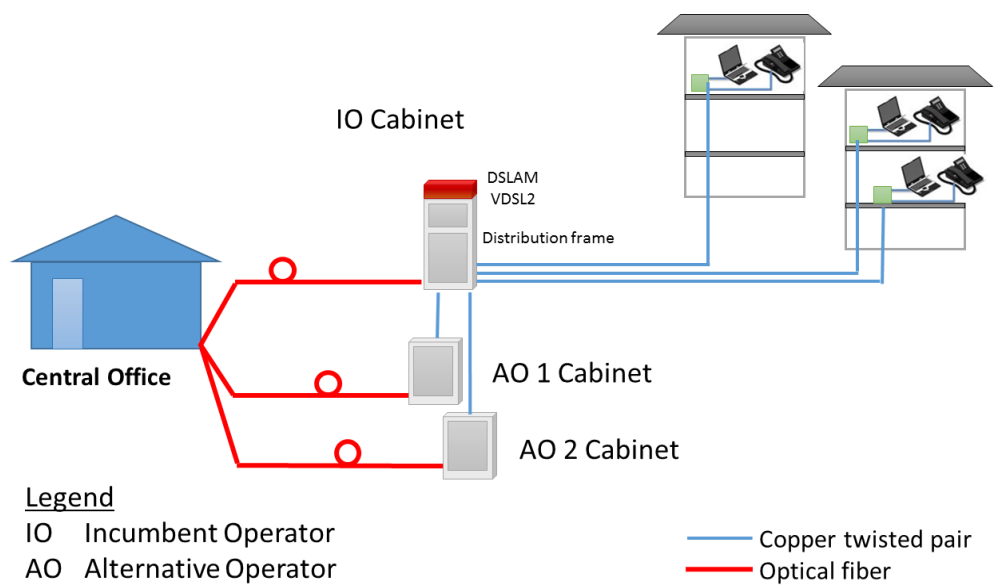

*Figure 2: Multi-CAB configuration (case of three operators).*

This regulatory regime encourages operators towards 30 Mbit/s commercial offers also favored by an increase in customers service demand. However, the regulation finds technical obstacles in promoting the coverage needed to meet the 100 Mbit/s objective for year 2020 of the Digital Agenda for Europe (DAE-2020).[3] In fact, while beneficial in promoting competition, the Multi-CAB solution prevents vectoring to be useful due to the uncompensated far-end crosstalk effect (FEXT) [8, 9].

In March 2015, the Italian Government approved a set of policy measures [10, 11], which includes € 6.5 bln State aids while also foreseeing investments of similar size in network infrastructures coming from private actors. The Italian strategy aims at bringing 100 Mbit/s coverage to 85% of the population by 2020, mostly with FttDp/B/H, as well as universal coverage of 30 Mbit/s with a blend of technologies including wireless. However, consent is not general among experts whether this bold, heavily subsidized approach may prove effective in such a short time.

This paper aims at introducing one possible solution to the problem of providing more than 100 Mbit/s coverage to a large population extent (70%, or so) in urbanized areas, useful for a Country such as Italy having all the severe constraints highlighted above. Of course, a solution to such a complex problem must entail an integrated set of measures, acting both at level of technology, regulation and State aid. We only concentrate on the regulatory framework and on the technological framework aiming at a low-cost solution mostly market-driven, which operators acting independently can implement in a short time. The presented approach could also be intended to complement a FttDp/B/H national plan such as [10] by easily solving local or regional unexpected difficulties.

As we verified with extensive computer calculations on the Italian IO's copper network, in a Multi-CAB architecture the current VDSL2 17a standard profile (having 17.6 MHz bandwidth) can be unable to provide the 100 Mbit/s speed required to meet the DAE-2020 policy target. In fact, speed is roughly limited to about 30 to 50 Mbit/s even in large cities where copper line lengths are among the shortest ones. To solve this problem, AGCOM proposed Multi Operator Vectoring (MOV) systems [12, 13]. However, these solutions are hardly feasible in a short time and without imposing large cost penalties to operators, having also to coordinate their deployments at every intended CAB. Hence, the need arises to seek new solutions to ensure more than 100 Mbit/s data-rates over copper networks, while preserving infrastructure competition at IO's street cabinets.

---

[3] As known, the ultra-broadband DAE-2020 objectives are two: more than 30 Mbit/s 100% population NGA coverage and more than 100 Mbit/s 50% high-speed Internet service adoption.

In this paper, we show how multiple operators can use frequencies higher than 17.6 MHz with the Sub-band Vectoring (SBV) technique we proposed [14, 15] to get 100 Mbit/s coverage for a large proportion of the target population. This technique is in line with some enlarged bandwidth new technological solutions presently under development (e-VDSL with 35.2 MHz bandwidth). However, such vendors' proposals are effective at increasing data-rate for one operator only, typically the IO. Therefore, this operator may provide network access to Alternative Operators (AOs) through active service-level remedies, such as the bitstream access or the virtual unbundled access, i.e. the so-called VULA (Virtual Unbundling of the Local Access).

The main purpose of this paper is to describe and discuss performance achievable with our proposed SBV solution in increasing the data-rate under multi-operator conditions, while the NRA safeguards and encourages infrastructure competition.

This paper structure is as follows. In Section 2, we examine the regulatory conditions able to promote both competition and UBB infrastructure deployment. We are especially interested in scenarios where infrastructure competition does not pre-exist in a Country. In Section 3, after having explained the limits in introducing vectoring techniques in a multi-operator environment, we describe the SBV solution to share bandwidth between two or more different operators at the CAB. In particular, we show that bandwidth enlargement to any extent turns out to be ineffective without SBV. Then, in Section 4, we use Italy as a case study to show how SBV can be effective in enhancing UBB coverage without penalizing infrastructure competition between operators. In order to provide quantitative results able to show the actual feasibility of the proposed technique in a real scenario, we evaluate the degree of coverage actually achievable for the Italian network in two clusters of cities defined by the Italian Government [10], having different degree of population density. Finally, we provide our conclusions in Section 5.

**2. Competition and infrastructure investments in regulated environments**

Since the inception of the telecom market liberalization, the Ladder of Investment (LoI) proposition provides the common cardinal regulatory framework in the EU, though NRAs generally put it into effect differently at Member State (MS) level. In particular, very variegated local national conditions across Europe forced each NRA adopting a different regulatory toolbox to promote NGA.

As is known, the LoI proposition states that a NRA imposes the IO a range of remedies to progressively encourage AOs to invest, so that they can climb from simple resale of products through service competition (static efficiency) up to infrastructure competition (dynamic efficiency) [16]. Therefore, the basic aim of the LoI approach is pursuing infrastructure competition, being it considered the most effective way to achieve consumer's welfare in the long-term. In spite of that, a NRA generally balances efforts to pursue a judicious compromise with short-term retail price levels by also promoting an appropriate degree of service competition.

The application of the LoI proposition, yet a delicate issue in the presence of a technologically stable copper network, is much more challenging when the network structure sensibly changes as is in the case of the emergence of NGA. The relationship between degree of regulation and some related topics such as the speed of NGA infrastructures development and the broadband/ultra-broadband adoption in Europe has been a strongly debated issue for several years.

Initially, when NGA in Europe was still at an early stage, researchers focused on data analysis of broadband penetration. Bouckaert *et al.* processed data collected for twenty OECD Countries between December 2003 and May 2008. Based on their study, they argued that [3]:

- Broadband penetration tends to be higher by about 12% in a Country where DSL and CaTV have equal market share, compared to a Country without cable operator(s).
- Intra-platform service competition based on ULL (Unbundling of the Local Loop) does not provide a significant effect on broadband penetration.
- Service-based competition within a platform, such as bitstream access, exhibits a significant negative effect on broadband penetration.

The above results point out that competition between different independent platforms is important for the deployment of broadband services. A larger market share of service-based competitors acting on a single platform is associated with lower rates of broadband penetration. Said differently, service-based competition could be more a barrier to the spread of broadband services, rather than an enabler. Therefore, policies that promote access to an IO's DSL network can adversely affect incentives to invest in the development of the network, especially if these policies are limited to promoting competition on services within the same platform.

More recently, econometric data started to be available able to show the relationship between infrastructure development and NGA regulatory policies in Europe. Authors of [4, 5] show that the scope and effectiveness of the regulation of wholesale broadband access adversely affect the adoption of the NGA. The impact of regulation turned out to be quite substantial. Consequently, said authors judge the DAE-2020's policy objectives to be inconsistent with the relevant EU regulation, which extends cost-orientation to access the new NGA infrastructures and the associated wholesale services. In addition, achievement by the year 2020 of the 30 Mbit/s coverage and 100 Mbit/s adoption objectives becomes much more difficult if rollout scenarios based on high-cost FttDp/B/H architectures are attributed primary importance. According to these studies, the impact of the effect of escape from competition surpasses the Schumpeterian effect. The inertia to remain on the 'old' network seems to be of particular relevance in some of the founding EU Countries with well-established infrastructures.

One main conclusion is that infrastructure competition highlights a positive impact both on networks investment and on adoption of UBB services. Service-based competition, conversely, correlates negatively with the investments, while the impact on ultra-broadband services adoption turned out to be less clear due to the uncertainty in assessing the traction effect of services, especially high-quality video services.

Within this picture, it is useful to revisit how the LoI proposition works in a scenario intending to promote efficient migration to the NGA. Practical experience has highlighted virtues, but also limitations in LoI's application, as it has been actually implemented [17]. In practice, we can identify three main difficulties from the point of view of NGA promotion:

1. European regulatory practice simultaneously put together an array of different remedies, and they generally stratify over time with ample differences from MS to MS.
2. In the presence of significant technological innovations, such as the introduction of optical technologies in the access network, one can also see the migration problem in terms of "two ladders put in parallel". The migration from one network (copper) to the other one (optical fiber) may exacerbate the difficulties of transition to dynamic efficiency [18].
3. The "rungs" of the ladders are not equally spaced in terms of economical efforts. As AOs climb with relative ease up to the ULL level (and, for a few of them, to the SLU), the ultimate step to full infrastructure separation implies complexity, costs and risks that may be highly discouraging.

Below we discuss some main issues and implications of above-mentioned difficulties in a LoI regulatory scenario.

**2.1. Array of regulatory remedies**

Although Cave's original formulation [16] advise that a Regulator "burns" over time the lower rungs not to let operators to step back on the ladder, this was not the approach adopted by the European Regulatory Group (ERG, now BEREC). The implementation of the LoI in Europe sometimes is considered one cause for emergence of the replacement effect [19] discouraging infrastructure competition.

The data collected so far seem to confirm that. As early as in 2007 Gruber [20] evidenced growing divergence between increasing adoption of DSL lines and investments in fixed lines decaying between 2002 and 2005: in particular, AOs appeared to display a poor level of investment. Subsequently, other authors [17, 21, 22] suggested that both IOs and AOs manifest little propensity in investing in infrastructures, both traditional and innovative ones, in a regulatory framework that encourages competition on services, including ULL. The choice not to burn regulatory rungs, while the array of remedies may grow up and stratify over time, is very likely to be one explanation for investment incentives for NGA in some European Countries not to increase at desirable pace.

Some studies predict very limited NGA coverage in Europe by AOs in the long run: in Countries where extent of alternative CaTV networks is limited (e.g., France, 10%) or null (Greece, Italy) the weakness of infrastructure competition coming from AOs only is one main reason for these MS lagging behind in NGA coverage. Analysys Mason provided one of the few available projections of NGA coverage level that allows to separately estimating coverage by IOs and AOs in Western Europe [23]. It appears that IOs will afford most of the NGA coverage (Table 1). Taking into account that part of it will be attributable to CaTV operators, other AOs will contribute few percentage points overall. As AOs will essentially compete at service level, the European body of *ex ante* regulation, as implemented today, does not appear to be very effective in incentivizing AOs to infrastructure.

| Country | UK | NL | DE | SE | ES | FR | IT | WE |
|---|---|---|---|---|---|---|---|---|
| **IOs** | 90 % | 88 % | 79 % | 66 % | 56 % | 53 % | 50 % | 68 % |
| **Others** | 6 % | 11 % | 6 % | 0 % | 14 % | 15 % | 3 % | 10 % |
| **Total** | 96% | 99% | 85% | 66% | 70% | 68% | 53% | 78% |

*Table 1: Forecast to E2018 of NGA coverage by incumbent operators and others (AOs including CaTV operators) in some European Countries and Western Europe average (Source: Derived from Analysys Mason data, 2013).*

In Italy, no CaTV network is present. Therefore, competition between telecom operators is the only form of competition in place. Italy is an interesting case study with respect to the progression of AOs along the LoI and the capability of a regulatory framework to stimulate infrastructure competition. While in 2008 new central offices (COs) open to ULL where 129 (out of the 10,500 available in the Country) in 2013 the yearly increment was only 20; a similar trend is evident on the number of new lines requested by AOs (net increment collapsed to less than thousand in 2013). In the last years ULL entered a stationary phase in Italy (Figure 3), while churn between operators happen without opening new COs to the ULL service. Meanwhile, NGA lines always remained negligible until 2012.

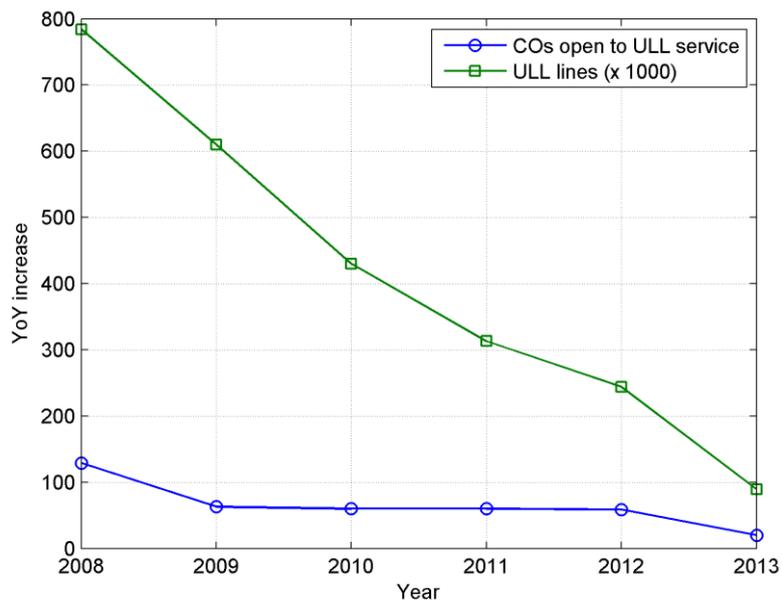

*Figure 3: YoY increments of ULL in Italy: Increments of number of IO's Central Offices available for ULL, and net increments of number of lines (in thousands).*

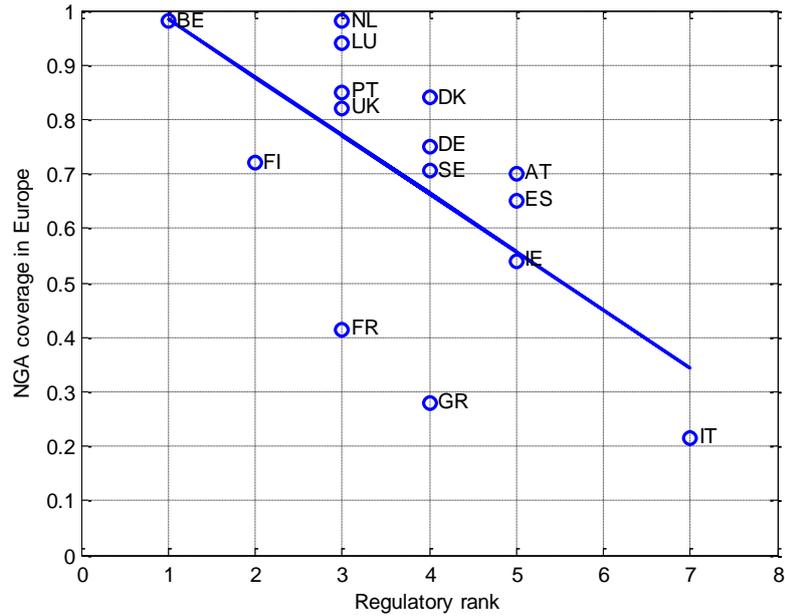

*Figure 4: NGA coverage in Europe vs. regulatory rank for several MS (Sources: for enumeration of remedies, Cullen 2015 data; for NGA coverage, DESI 2015).*

If the NRA does not "burn" nor weaken regulatory remedies over time, they tend to accumulate, so representing one reason for market stagnation and discouraging investment in NGA infrastructures. A simple picture can display a certain degree of correlation between burdensomeness of remedies and NGA coverage degree accomplished across EU Countries. In Figure 4, we use a parameter, which we call the regulatory rank, defined as the number of remedies imposed to the IO access network in a certain Country, and relate it to the NGA coverage provided at the same date. Even such an extremely simplified view, which neglects the relative severity of remedies, can evidence a clear trend towards less NGA coverage as the number of remedies increases in a Country.

**2.2. Two ladders in parallel**

Worldwide we can find few experiences of rapid and direct transformation of a national access network into full optical access. Some prominent examples of total replacement are rooted into poor conditions of pre-existing copper voice networks, unable to upgrade quality of service to the level needed for broadband access (e.g., some Eastern European Countries). In a few specific cases, political instances drove decisions, however generally with limited success until today (e.g., Australia). Instead, most of the Countries decided following migration paths with overlay period of copper and optical technologies. Under such conditions, NGA cost reduction and pricing policy are central issues for the Regulator to incentivize migration.

According to Cave [18], a metaphor of "two ladders" can describe migration scenarios: the first ladder refers to the legacy copper network while the second ladder is the NGA. Having both ladders in front, an AO could decide to go up along the new NGA ladder by: (a) requesting the IO subloop unbundling and rental of the final copper tract or, alternatively, (b) building or renting a duct to reach customers directly or, finally, (c) migrating to a higher capacity bitstream active access product. In most cases,

"lateral stepping" to the NGA ladder may be risky, as climbing up along the second ladder entails affording larger and uneven steps from FttC to FttDp and, then, to the ultimate FttH architecture.

To stimulate the climbing process and affording lateral jumping on the second ladder, one main lever in Regulator's hands is tariff policy. The approach selected by the EU was to identify a balanced regulation able to blend nondiscriminatory rules with costing and tariff incentives. One main element is how to fix wholesale tariffs for the legacy copper network services.

One simple rationale for tariffs on copper is that, if prices for the different rungs on the first ladder are 'high', jumping on the second ladder can be an option for the AOs. However, this may be an oversimplification and fixing regulated prices is a hard task because several intertwined factors intervene. If prices are too high, AOs may not have sufficient resources to invest and step up the ladder; conversely, if prices are too low they do not have sufficient incentive. In the latter case, an AO may acquire customers with relative ease, and revenues earned through service competition is an opportunity cost for infrastructure competition that generates the "substitution effect".

As seen from the IO side, if prices are 'low' financial resources may lack as willingness to invest in risky new technologies may be insufficient in a regulated sector with low expected WACC and uncertain returns profile. However, this is not the only effect. For the IO, the incentive to invest in a 'new' infrastructure and the profitability of the services provided on the 'old' one are related factors. A higher access price to the 'old' network increases the opportunity cost, and this "wholesale revenue effect" tends to disincentive the IO to invest in a network of better quality, as the competitors may react by investing in turn, so that the IO will lose to some extent wholesale profits. Additionally, if prices are low and one form of functional separation is applied inside the IO's organization (e.g., Italy, UK) between a "wholesale division" and a "retail division" strictly separated, the former tends increase clients (the AOs) while the latter loses theirs (the final customers). The interest of the IO's wholesale division converge with that of AOs, so the substitution effect and the wholesale revenue effect mutually reinforce. This increases the barrier to the infrastructures creation.

We must also consider the customer's point of view. When the access prices on the copper network are low, retail prices for services that rely on this network are low, therefore, in order to encourage customers to switch from the 'old' network to the 'new' network operators should also offer low prices on the latter. This "business migration effect" reduces the profitability of the new technological infrastructure and thus the incentives to invest in it.

Finally, from a general perspective, prices should also converge at European level, in order to avoid that investors tend to skip those Countries where wholesale prices are relatively low, so enlarging the fork between Countries in terms of coverage of the new infrastructure.

Following a long and heated debate, in 2013 the EC has welcomed the result of a study by Charles River Associates. It took jointly into account the replacement effect, the wholesale revenue effect and the business migration effect to conclude that ULL prices on copper networks need to increase and tend to harmonize at European level as a necessary tool to stimulate NGA networks and to speed up the migration of customers to the optical network [24].

Although pricing policies are important, they may be insufficient in some scenarios, especially where infrastructure competition in chronically low and the demand push is insufficient. In such cases, the virtues of the LoI could be important. However, they can collide against cost barriers. As seen above, fine-tuning of regulated prices is a very complex task in a NGA scenario, with so many variables interacting that the result of a pricing policy can be disappointing. Therefore, we now turn our attention on how to moderate costs for the benefit of both the IO and the AOs.

**2.3 Uneven spacing between rungs on a ladder**

When CaTV networks are well developed, NRA's effort is facilitated. Generally, CaTV providers are not classified as dominant telecom operators (as geographic markets are little used across Europe): therefore, they are out of interest of NRAs and only subject to *ex post* antitrust rules. In addition, NGA competition is already in place and the NRA must only impose few, and light, remedies on the IO. The LoI proposition is still valid but it does not exert in full its expected benefits, and telecom AOs tend to remain weak. In some of these Countries, the NRA does not impose the SLU, so that the IO can offer VDSL2/FttC with vectoring. Therefore, customers have good chances to achieve 100 Mbit/s, or more, data-rates from either the IO or the local CaTV EuroDOCSIS 3.0 offer.

Contrarily, in those Countries where cable TVs are limited or null the LoI could prove beneficial. However, as we saw above, in some cases as in Italy due to limited demand, along with a complex regulatory scenario, LoI's advantages hardly manifest themselves in full. Especially in such scenarios, one main problem apparently not discussed enough in the literature is the uneven separation between ladder's rungs. Stepping up one level on the first ladder can be feasible, when there are sufficient signals coming from the demand side. However, jumping over the second ladder can be one order of magnitude more difficult, especially if demand is weak. Certainly, the direct jump to FttDp/B/H presents risks and costs often difficult to afford, in spite of low-cost availability of ducts and other civic infrastructures mandated to municipalities. Therefore, ascending the ladder directly to FttH (or FttDp) and, in certain conditions, even to FttC could be too much a challenge for operators, when demand signals are insufficient.

There is a lack of technological tools, which could provide additional ladder rungs to reduce costs and distribute them over time. However, one problem is standardization. Standards are provided at international level and they generally require a very long production cycle (even more than five years), which can be incompatible with the need for appropriate technological tools. Therefore, the only way to provide effective technological solutions, which can be standardized in a short time (e.g., one year), is to enhance standard capabilities through limited modifications. One such solution, which some vendors are pursuing, intends to provide extended bandwidth to DSL standards. As is intuitive, a larger bandwidth can provide increased data-rate in a FttC (or FttDp) scenario, possibly allowing the increase of 100 Mbit/s NGA coverage especially in Countries where the CAB median distance from users' premises is short (e.g., 200-250 *m* as in Italy). However, in a competitive scenario the situation is more complex as the above intuitive picture suggests. This is what we are going to discuss in the following Section.

## 3. Enlarged bandwidth DSL solutions

### 3.1 The intricacies of vectoring in a competitive environment

Especially when operators use very large bandwidths on copper pairs and for long distances, FEXT and attenuation limit the maximum achievable data-rate. To counter the effects of FEXT and substantially improve the performance of VDSL2 and G.fast (Fast Access to Subscriber Terminals) vectoring can be used, a transmission technique originally proposed by Ginis and Cioffi [23] in 2002, and then standardized as ITU-T G.993.5. Vectoring ideally eliminates FEXT in the downstream direction (DS) to the customer through interference pre-compensation, as well as in the upstream direction (US) by its cancellation.

Since both the FEXT intensity and the attenuation rapidly grow with frequency, in FttC practical conditions the copper pair's bandwidth may be useless above 17.6 MHz, which is the upper frequency of the VDSL2 17a spectral profile, the one most used in Europe. Even for frequencies below 17.6 MHz, FEXT may produce a significant performance degradation, which reflects on a severe reduction in the data-rate otherwise attainable on an ideal AWGN (Additive White Gaussian Noise) channel.

However, when activated by a single operator that controls all copper pairs present in the cable, vectoring allows the use of frequencies well above 17.6 MHz. At present state-of-the-art, the maximum frequency can extend to 200 MHz, and possibly in the future even more. Thanks to vectoring, under such conditions the cable channel ideally turns out to be AWGN. Unfortunately, even very few copper pairs not centrally controlled, especially if belonging to the same cable's binder, may lead to a very marked degradation of transmission quality on the group of vectored lines [8, 9]. In fact, uncontrolled copper wires originate the so-called "alien-FEXT", that is the cause of severe performance degradation on the vectored group. A special case, but very relevant in practice, is when two (or more) operators independently implement vectoring on the same cable, each one controlling separately one vectoring group only. Therefore, each vectoring group acts on other ones as alien-FEXT. The above makes it difficult to reconcile the effective use of vectoring where the NRA imposes SLU at an incumbent's CAB or, equivalently, Distribution point (Dp).

The controlled conditions of operation making it possible to combine vectoring and SLU are difficult to achieve. To this aim, MOV systems have been proposed [13] being incompatible with already installed technologies and systems, and that face many obstacles. Some obstacles are that they:

- Require an international standard for ensuring interoperability between different vendors equipment in order to achieve appropriate economies of scale;
- Make use of complex equipment not always easy to install;
- Might suffer from privacy issues between different operators and difficulties in ensuring the confidentiality of customers information;
- Must be powered on, and energy consumption may not be negligible, so that the operating cost for the telecom operators increases;
- May have significant impact on the ecological footprint.

Given the MOV difficulties and limitations, some vendors proposed solutions for enlarging the bandwidth of the copper pair to obtain an increased data-rate in the presence of vectoring at a CAB. Today's state-of-the-art for the DSL access multiplexer (DSLAM) bandwidth larger than that of VDSL2 17a profile foresees solutions with 35.2 MHz band. In fact, a few major vendors started the process of international standardization based on commercial solutions little differentiated (e.g., see "Vplus" [25] and "SuperVector" [26]). Collectively, e-VDSL (Extended-bandwidth VDSL) references those commercial proposals. Such solutions are promising, also by virtue of the possible use of powerful channel codes, so that the reach extension for one single operator can be well beyond the typical lengths of VDSL2 with vectoring (e.g., 400 Mbit/s within 300 $m$ and 100 Mbit/s within 800 $m$, according to [26]). However, they are not able to solve the problem of coexistence of two or more co-located operators as the alien-FEXT, in the presence of distinct vectoring groups, destroys the benefit of bandwidth enlargement, as we will show later on in the paper.

To sum up, the problem at hand appears intricate. In some Countries, the most advisable solution to promote infrastructure competition is to retain the SLU obligation at a CAB or Dp. However, without vectoring the performance attainable is poor. On the one hand, very complex MOV systems appear far from an implementation while, on the other hand, also larger bandwidth DSLAM do not help in the presence of alien-FEXT.

To solve the problem, we introduced a new family of techniques under the common denomination of Sub-band Vectoring (SBV), based on frequency division multiplexing and that allows vectoring use without sacrificing SLU [14, 15]. The SBV can be implemented in a short time without complex coordination nor synchronization between co-sited operators.

### 3.2    Sub-band Vectoring principle and performance

SBV denotes a new family of system solutions suitable to achieve the ultra-broadband DAE-2020 objectives. It is based on a two-level spectrum partitioning. The first level of partitioning in sub-channels is performed in order to preserve backwards compatibility with VDSL2 profile 17a use, while extending the service bandwidth beyond the 17.6 MHz limit. By so doing, possible pre-existing regulation for usage of the lower 17.6 MHz band is not affected. The second level of partitioning refers to the higher band, up to 35.2 MHz for e-VDSL or even more for possible future standards. The higher band is divided in smaller sub-bands (e.g., 5 MHz blocks), and separately allocated to operators in order to ensure the respect of one fairness principle [14], which is essential to avoid that customers of different telecom operators served on the same cable experience different data-rates at the same distance from the CAB. Figure 5 shows the SBV concept with its two-level spectrum partitioning approach.

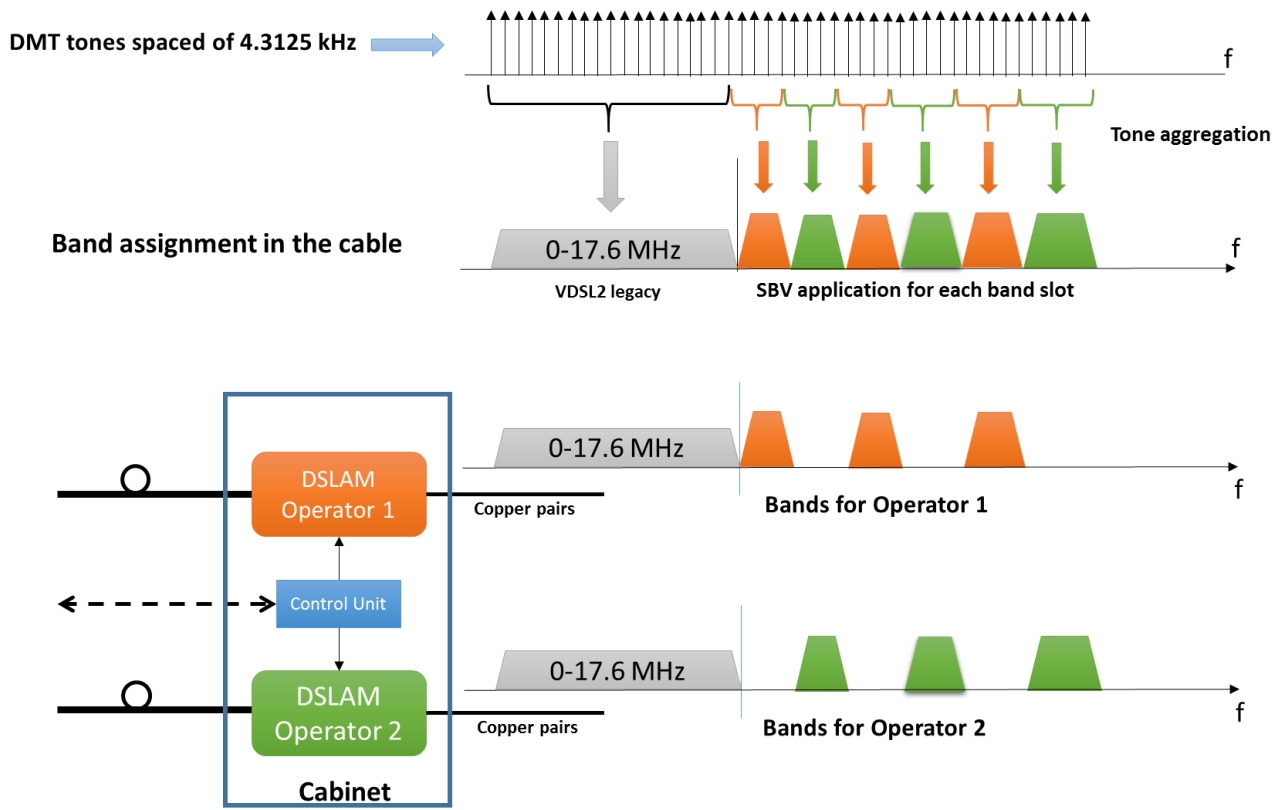

*Figure 5: Band partitioning in the SBV technique with division in sub-channels and in sub-bands, and rule of sub-bands allocation between operators (case of two operators).*

In a multi-operator environment, extended bandwidth solutions, such as the 35.2 MHz e-VDSL, ensure a real advantage when they are combined with SBV. In fact, as we showed elsewhere [14], alien-FEXT dramatically destroys the benefit of the spectrum expansion beyond a frequency in the range 10-15 MHz depending on several operational conditions. This is one reason why, in practice, SBV adopts vectoring in disjoint bands above 17.6 MHz, while it retains non-vectoring (NV) below this frequency threshold.

To appreciate the severity of the penalty that alien-FEXT, along with attenuation, procures above 17.6 MHz without SBV consider the results of the computer calculations shown in Figure 6. Our calculations take as a reference the VDSL2 profile 17a band-plan up to 17.6 MHz, assuming for simplicity the additional spectrum beyond this frequency being allocated to DS only. We assume the standard cable LQ-Gamma model valid up to 200 MHz.[4] Performance are considered at a distance of the CAB from the subscriber d = 100 *m*. We reported the cases corresponding to a number of operators $N_{op}$ =2, 3 and a number of interferers $N_{us}$ = 12 (medium traffic load), 24 (high traffic load). To account for the (un-avoidable) presence of residual FEXT after vectoring due to non-idealities (fixed point algebra, channel estimation errors, non-ideal channel matrix inversion, etc.), we assume an additional degradation $r_v$ =10 dB due to the imperfections in the vectoring algorithm implementation expressing the residual FEXT as an increase of background noise.

---

[4] The LQ-Gamma model in the 0-17.6 MHz bandwidth has similar performance of the AWG 24 cable, which represents the most frequent choice in the Italian access network. However, to the best of these authors' knowledge there is no model characterizing the AWG-24 cable for the considered frequency range, i.e. up to 100 MHz. This is the reason why we only refer to LQ-Gamma in this paper.

Figure 6 compares the values of data-rate when SBV is adopted and when it is absent, i.e. when the entire available spectrum is shared by operators (non-vectoring case, NV). It confirms that the bandwidth enlargement is ineffective in the presence of vectoring if alien-FEXT is not cancelled while, at the same time, the number of interfering lines, $N_{us}$, has little influence in case of SBV. For SBV, variations with $N_{us}$, are due to the sharing of the 0-17.6 MHz (non-vectored) band among operators.

In conclusion, while the alien-FEXT is significant enough to cancel in practice any benefit of the increased bandwidth, ideally each channel becomes AWGN if SBV is adopted. In the latter case, among other benefits, powerful channel codes can be effective and can produce high values of coding gain. Therefore, the combination of the SBV technique and channel coding may be one key to resolving the UBB coverage problem of a Country with multiple competing operators, at modest infrastructure incremental costs.

In order to estimate this result quantitatively in terms of achievable coverage, we now turn to examination of the possible application of the SBV technique in the case of the Italian network.

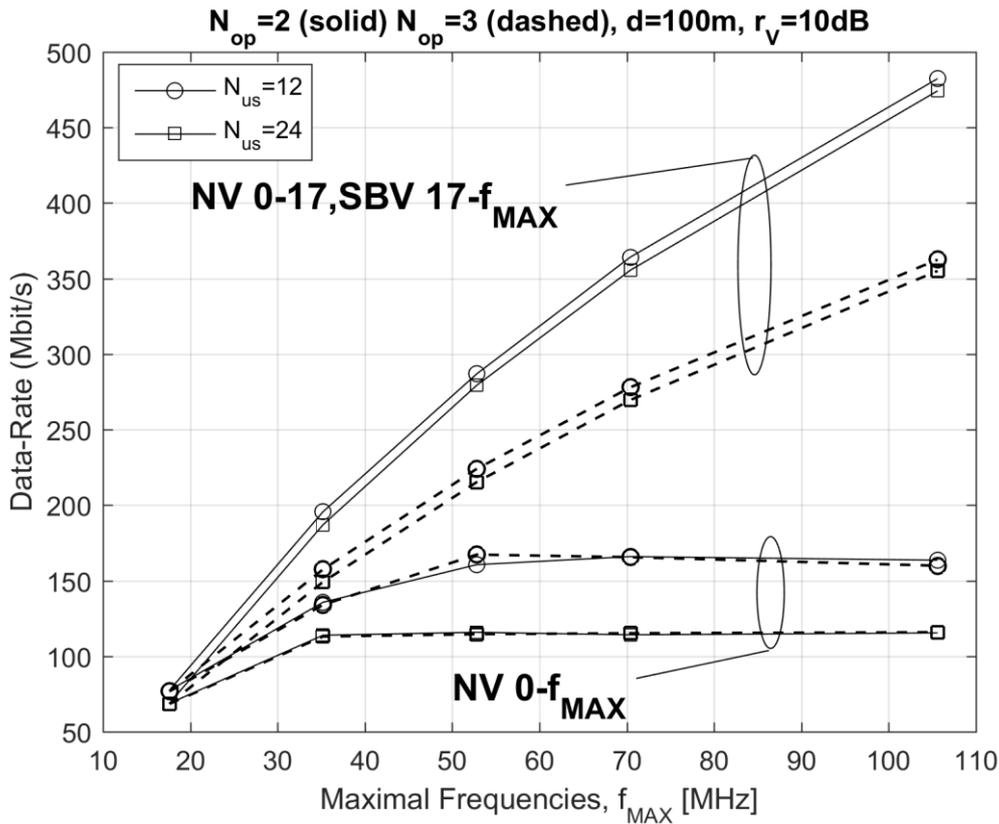

*Figure 6: Data-rate (Mbit/s) per operator achievable in the presence of SBV and in its absence (NV) as a function of the DSLAM bandwidth.*

## 4. Italy as a case study

In order to better target scarce economical resources for public investments, the Italian Government classified the territory into "Clusters" [10] taking into account several factors among which: strength of demand, level of achieved penetration of broadband services, population distribution, geographical characteristics, etc. Therefore, it identified four clusters, implying different costs and complexity for infrastructure implementation. These clusters are as follows:

- Cluster A (i.e., "black area"): it consists of the 15 most populated cities totaling 9.4 million inhabitants (i.e., 15% of residential population), where at least two telecom operators expressed interest to invest in order to provide 30 Mbit/s (or higher speed) networks. Then, Cluster A is a competitive market area for operators, and it is the most attractive area for investments (no public intervention needed).
- Cluster B (i.e., "dark-grey area"): it consists of about 1,120 medium populated cities totaling about 27 million inhabitants (i.e., 45% of residential population), where at least one telecom operators deployed, or is going to deploy, 30 Mbit/s networks but operators are not presently interested in 100 Mbit/s, since return of investment is not ensured.[5]
- Cluster C (i.e., "light-grey area"): it consists of about 2,650 towns with about 14 million inhabitants (i.e., 24% of residential population). They are now generally covered at 2 Mbit/s and are market failure areas for ultra-broadband, i.e. they include areas where operators will not invest to deploy a 100 Mbit/s network, except possibly with public aids.
- Cluster D (i.e., "white area"): it consists of about 4,300 small towns in rural areas, with about 8 million inhabitants (i.e., 13% of residential population). It consists of the rest of market failure areas,[6] where operators will not invest to deploy a network at 30 Mbit/s, so that deployment of a public network is envisaged.

Table 2 summarizes the main characteristics of the Italian clusters.

| Cluster | Munici-palities | Population | | Households | Type | Investments | Target | |
|---|---|---|---|---|---|---|---|---|
| | | Million | % | Million | | | DS (Mbit/s) | US |
| **A** | 15 | 9.4 | 15 | 3.9 | Black | Private | 100 | N.D. |
| **B1** | ~ 1,120 | 27.0 | 45 | 11.2 | Grey | Private | 100 | |
| **B2** | | | | | Grey | Public | | |
| **C** | ~2,650 | 14.0 | 24 | 5.8 | Grey | Public | 100 | |
| **D** | ~4,300 | 8.0 | 13 | 3.4 | White | Direct | 30 | |
| N.D.: Not defined | | | | | | | | |

*Table 2: Italian clusters' characteristics.*

---

[5] Cluster B is further split into Sub-cluster B1, where private operators are investing, and Sub-cluster B2, where limited public investments are currently going on. In this cluster, public investments are mostly intended to upgrade network speed to 100 Mbit/s.
[6] A further 3% of the Italian territory consists of sparse homes and not inhabited areas.

### 4.1 Achievable Coverage in Clusters A and B

Thanks to favourable secondary access conditions, Italian scenario for ultra-broadband access offers interesting opportunities for rapid, effective and relatively low cost deployment of the e-VDSL technologies over a large portion of the national territory.

In this Section, we analyse the FttC/DSL system architecture including the SBV technique for DS transmission, taking as a reference Italian Clusters A and B. We will not consider Clusters C and D. Differently from those, from a technological point of view Clusters A and B are suitable to massively deploy a 100 Mbit/s network. This can be done efficiently and economically by promoting Multi-CAB infrastructure competition conditions. Therefore, we now consider a network scenario useful to estimate achievable coverage levels for Clusters A and B in Italy.

Performance presented in this Section is expressed in terms of the achievable coverage in Clusters A and B. Considering the scenario introduced in the previous Section and detailed in [14], coverage is obtained from curves of data-rate vs distance and the realistic distribution of the distances of the user terminals from CAB on the Italian secondary network (see Figure 1).

The overall CAB-to-terminal distance of the generic reference terminal is given by the sum of the CAB-to-Dp distance with the Dp-to-home distance. Distance is then used to evaluate the achievable data rate from the results provided in [14] (see before) and it is subsequently stored to evaluate the coverage, i.e. the probability that the achievable data-rate is greater than a given threshold.

As in [14], we assume the band above 17.6 MHz is used for DS transmissions only. For the SBV case, the overall (average) aggregate DS data-rate for the reference DSL terminal is the sum of the data-rate achieved using the standard non-vectored VDSL2 17a profile in the 0-17.6 MHz band with the data-rate obtained in the higher bands (up to 105.6 MHz).

### 4.2 Results and comments

Having shown above that under NV conditions the bandwidth enlargement is not useful, we first show how SBV turns out to be effective as bandwidth extends from $f_{MAX} = 32.5$ MHz (e-VDSL case) to $f_{MAX} = 105.6$ MHz. Figure 7 shows the complementary cumulative distribution function (C-CDF) per each of three operators in Cluster A, comparing NV and SBV conditions. C-CDF expresses the achievable coverage. Results are shown for different values of $f_{MAX}$ with 24 interfering pairs in the same cable binder and accounting for 10 dB vectoring degradation in case of SBV. While, as expected under NV conditions, the DAE-2020 100 Mbit/s target can be achieved only for small percentages (20%, or so) when SBV is adopted this target is met for a fraction of population between 60% and 70%. It is worthwhile to note that, with larger bandwidth systems, for appreciable percentages of customers the data-rates of 150-200 Mbit/s are reasonable targets.
In the following figures, we concentrate on e-VDSL performance with and without SBV. Let us first consider Cluster A with three operators present at every possible CAB: this is a limiting case in terms of competition (worst-case in terms of coverage).

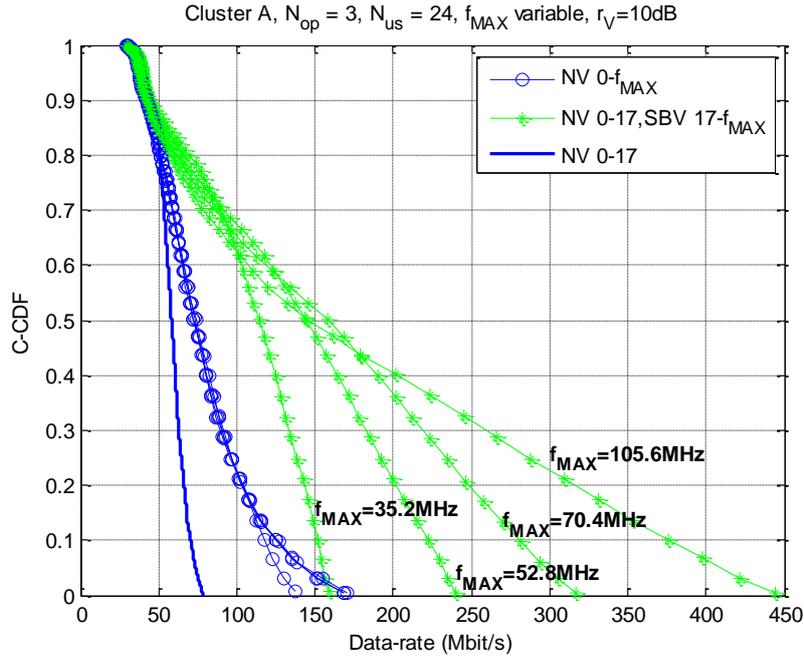

*Figure 7: Cluster A coverage (C-CDF) for one out of three operators as a function of data-rate (Mbit/s) assuming maximum frequency $f_{MAX}$ as a parameter, in case of NV and SBV, respectively.*

In Figure 8 we assume 24 interfering pairs (case a) and 12 interfering pairs (case b), respectively. With the assumed degradation due to vectoring imperfections, even under maximum interfering load the 100 Mbit/s target is achieved for more than 60% coverage. In practice, the expected coverage could be higher as the assumed telecom operators will not be present at all CABs, the interfering load could be different from maximum, and when customers increase beyond a certain threshold, it is economically viable for operators to densify their CABs and/or Dps.

For comparison purposes let us consider the situation shown in Figure 9, which replicates all the conditions already present in Figure 8, in the case of $N_{op}=2$. As expected, the coverage with SBV is now between 70% and 75%.

Being Cluster B less competitive, let us consider now only two operators at every CAB. Coverage results are shown in Figure 10. Under the same conditions that led to the results for Cluster A, we see that 100 Mbit/s coverage is now ensured for 60% to 65% of the population.

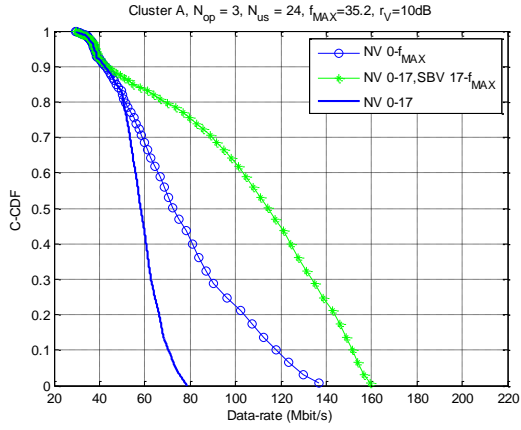
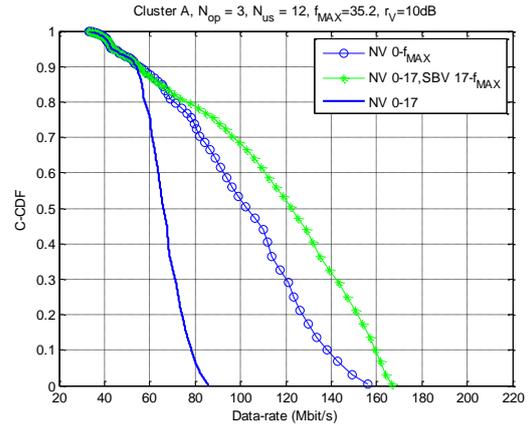

*(a)* *(b)*

*Figure 8: Cluster A coverage (C-CDF) for one out of three operators as a function of aggregate data-rate (Mbit/s) for e-VDSL in case of NV and SBV, respectively; coverage is also compared with the case of VDSL2 profile 17a: (a) 24 interferers; (b) 12 interferers.*

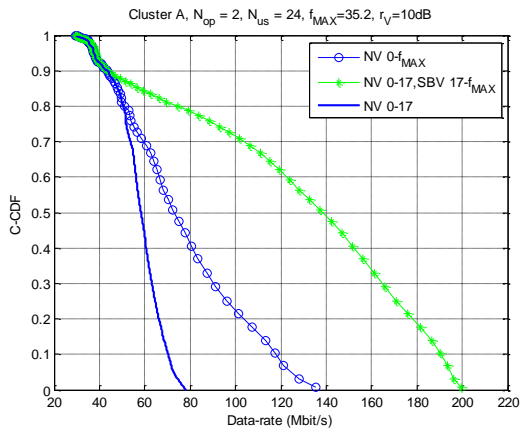
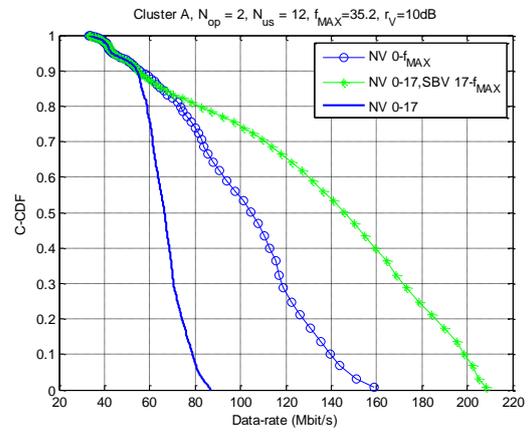

*(a)* *(b)*

*Figure 9: Cluster A coverage (C-CDF) for one out of two operators as a function of aggregate data-rate (Mbit/s) for e-VDSL in case of NV and SBV, respectively; coverage is also compared with the case of VDSL2 profile 17a: (a) 24 interferers; (b) 12 interferers.*

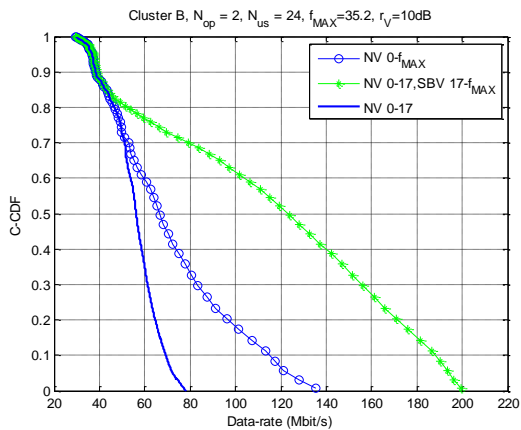
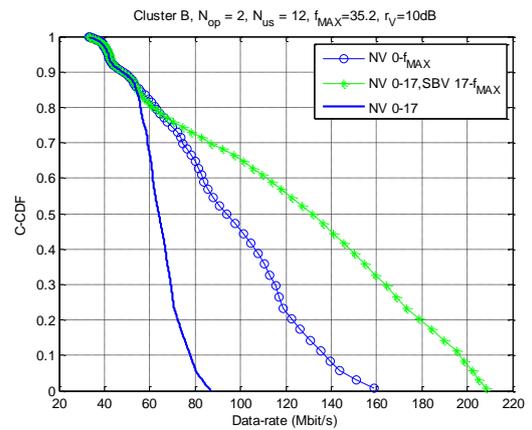

*(a)* *(b)*

*Figure 10: Cluster B coverage (C-CDF) for one out of two operators as a function of aggregate data-rate (Mbit/s) for e-VDSL in case of NV and SBV, respectively; coverage is also compared with the case of VDSL2 profile 17a: (a) 24 interferers; (b) 12 interferers.*

## 5. Conclusions

This paper aimed at introducing and discussing the SBV (Sub-band Vectoring) technique for sharing band among telecom operators independently implementing vectoring in a FttC (or FttDp) architecture. To begin with, we examine the regulatory conditions able to promote both competition and UBB infrastructure deployment. As the most effective way to promote NGA coverage and UBB penetration is through infrastructure competition, we try ascertain how this condition can be better surrogated in Countries, such as Italy, where it does not pre-exist. In Italy, to promote the 30 Mbit/s DAE-2020 objective, the NRA selected an approach hinged on SLU regulation and Multi-CAB architecture, which is manifesting virtuous effects. However, some technical intricacies make hard using vectoring. Therefore, our SBV proposal along with the new e-VDSL technology turned out to be a solution able to boost the data-rate well beyond 100 Mbit/s. In fact, SBV allows the simultaneous use of UBB data transmission of co-sited operators not coordinated nor synchronized with each other. It is a procompetitive, fair and neutral solution. It is also simple, and it can be implemented in a short time.

In conclusion, we believe that, thanks to its virtues, SBV is an interesting candidate to fill an evident technological and regulatory gap to accelerate the migration path from copper to all-fiber networks in Europe and elsewhere.

**Acknowledgements**
Work sponsored by KRUPTER SrL, Rome, Italy